\documentclass[preprint,12pt]{elsarticle}

\usepackage{amssymb}
\usepackage{amsthm}
\usepackage{amsmath}
\usepackage{tabularx}
\usepackage{multirow, hhline}
\usepackage{pgfplots}
\pgfplotsset{compat=1.5}
\usepackage{url}

\journal{Digital Signal Processing}

\begin{document}

\begin{frontmatter}



\title{Active Voice Authentication}


\author{Zhong Meng, M Umair Bin Altaf, and Biing-Hwang (Fred) Juang
}
\address{Georgia Institute of Technology, Atlanta, GA 30308, USA}

\begin{abstract}
 Active authentication refers to a new mode of identity verification in
  which biometric indicators are \emph{continuously} tested to provide
  real-time or near real-time monitoring of an authorized access to a
  service or use of a device. This is in contrast to the conventional
  authentication systems where a single test in form of a verification
  token such as a password is performed. In active voice authentication
  (AVA), voice is the biometric modality. 
  This paper describes an ensemble of techniques that make
  reliable speaker verification possible using unconventionally short
  voice test signals.  These techniques include model adaptation and
  minimum verification error (MVE) training that are tailored for the
  extremely short training and testing requirements. A database of 25
  speakers is recorded for developing this system. In our off-line
  evaluation on this dataset, the system achieves an average
  windowed-based equal error rates of 3-4\% depending on the model
  configuration, which is remarkable considering that only 1 second of
  voice data is used to make every single authentication decision. On the NIST SRE 2001 Dataset, the system provides a
  3.88\% absolute gain over i-vector when the duration of test segment is
  1 second. A real-time demonstration system has been implemented on
  Microsoft Surface Pro.
\end{abstract}




\end{frontmatter}






\section{Introduction}
User authentication refers to the process of validating 
a user's claim of identity in order to grant or deny the user access to 
a device or service. The prevalent method for user authentication operates 
in predominantly the so-called gatekeeper mode, in that the guarding system 
asks the user to present what he/she knows (e.g., a password), what
he/she has 
(e.g., a key or a fob), or what he/she is (e.g., fingerprints, iris scan) for 
examination in order to render the decision. Once the access is granted, 
the device or service remains ``active'' until it is signed off or terminated. 
During the active session, no action is taken by the guarding system even 
though the user may have changed, resulting in security compromises.

An active authentication (AA) system seeks to actively
and continuously validate the identity of the person by making use of
his or her unique biometric identifiers without repetitively prompting the user
for credentials or requiring the user to change his/her work-flow during
the use of the device or service. The
AA framework differs from a conventional authentication system in
that it provides a continuous and real-time monitoring of the user's
identity rather than just a one-shot authentication in form of
verifying a test token as in the gatekeeper mode.

Many biometric identifiers including physiological and behavioral
indicators can be used as human characteristics to
actively verify the identity of a user \cite{biometrics_define_acm}. As the intrinsic
attributes, the facial appearance \cite{face_acm}, the iris pattern \cite{iris_ieee}, the finger-print \cite{fingerprint_acm}, the voice
pattern \cite{voice}, the hand geometry \cite{hand} and body's electric pulse response \cite{body_pulse_response} are widely used as the physiological
identifiers while the manner people walk \cite{walk}, write \cite{write}, and
type \cite{wang2012user} are commonly used as the behavioral identifiers. 

The target modality in this paper is
the voice. The voice of a person is
unique. This is because the construction of the articulatory apparatus
and its use that generate and modulate the voice of a
talker\textemdash{}the lungs, the vocal cords, the articulators,
etc.\textemdash{}are uniquely configured for a given individual and this
configuration is naturally embedded in the person's voice
characteristics. Thus, in addition to language, voice conveys the latent
identity of its speaker. Human voice is ideally suited for AA as it
provides contact-less authentication; it is straightforward to acquire
authentication data from ubiquitous microphones available on all
platforms. An active voice authentication (AVA) system uses the voice of
a person to achieve AA, as the person uses the phone or any other voice
application on a mobile or desktop device. The AVA system
does not interfere with other active authentication methods on the device
and can work in the background with already-installed voice applications,
such as Skype, Voice note or the Phone to provide real-time continuous monitoring of the user's identity. AVA can effectively strengthen the security of the existing voice assistants such as Amazon Alexa, Microsoft Cortana, Apple Siri and Google Home \cite{hoy2018alexa, lopez2017alexa} and enable services which involve money transfers. On top of the initial speaker verification using the starting anchor (wake) word, AVA continues to repeatedly authenticate the user's voice as the conversation goes on and reserves the right to overturn its initial decision at any time. 

Recently, voice has been successfully used to assist the other biometrics such as body-surface vibrations \cite{feng2017continuous}, touch gestures \cite{peng2017continuous} and a combination of face, mouse and keystroke \cite{fenu2017multi} in performing continuous authentication. In these works, the voice is authenticated in form of voice commands which are first stored as audio files and are then verified through support vector machine (SVM) \cite{feng2017continuous, peng2017continuous} or vector quantization \cite{fenu2017multi}. Although these systems perform further authentications after the session is activated, voice only plays an \emph{auxiliary} role in protecting the system because the decisions are still made at utterance (command)-level as in conventional speaker verification with a low resolution of 3 seconds or more. The other biometrics are necessary for AA especially when the speaker is not talking. Therefore, these voice-assisted authentication systems do not meet the requirement of AVA.


As with any AA system, AVA involves two phases, the registration phase and the
authentication phase. During registration, the user being registered is asked to utter some standard speech material. A set of statistical models is trained a adapted to signify the user's voice identity. At active authentication stage, once AVA system detects a valid voice signal through a voice activity detector (VAD), it starts continuously evaluating real-time confidence scores given the speech signal. Depending on the score, the system can grant or deny the user's access to the device. If silence is detected to be long than the latency, AVA can report an authentication score that indicates impostor.

AVA is significantly different from traditional speaker verification
task directed and organized by NIST Speaker Recognition
Evaluation (SRE). The goal of AVA is to continuously
authenticate the speaker identity with the assumption that
change of talker can potentially occur at any instance whereas
in most SREs, such an abrupt change of talker does not happen and
its goal is to produce a final decision after the entire test
utterance is obtained. Because of the large distinction
between AVA and the traditional speaker verification, a new
design framework is necessary for the AVA system which we will elaborate in Section \ref{sec:real_time}.
The AVA system integrates the techniques of sequential
training and testing, maximum a posteriori (MAP) adaptation,
cohort selection and minimum verification error (MVE). The major contributions of this paper are the following:
\begin{itemize}
    \item Propose a novel AVA framework that continuously verifies the speaker's identity and instantaneously reports verification decisions.
    \item Propose a window-based short-time sequential testing scheme to accommodate the real-time requirement of AVA.
    \item Propose a window-based short-segment training scheme to model the short-time statistics of a speaker's voice through an HMM and to match the real-time testing condition.
    \item Apply MAP adaptation of an speaker-independent (SI) HMM to minimize the enrollment data needed for reliable short-time speaker modeling.
    \item Apply MVE training to further minimize the speaker verification error on top of MAP. Propose cohort selection method to address the imbalanced target and impostor data for MVE training.
\end{itemize}

AVA
performs speaker verification using second-long speech signals
and achieves a performance of 3-4\% average window-based
equal error rate (WEER), depending on the model configuration.
This level of performance, being able to reasonably
authenticate a talker's claimed identity with 1 second voice,
outperforms conventional techniques, as will be reported in
later sections, and outstrips human capabilities based on
the informal observation of our research group
members. A separate talker authentication evaluation on
human performance is necessary to formally establish the
comparison. We also evaluate the proposed methods on NIST SRE 2001 dataset with a large number of speakers and  the proposed system provides a 3.88\% absolute gain over i-vector on the  when the duration of test segment is 1 second.

The rest of the paper is organized as follows. 
In Section \ref{sec:principle}, we briefly discuss the conventional formulation of the problem of
speaker identification and verification. We explain how the differences between AVA and
the traditional talker verification paradigm would call for a
new design methodology. In Section \ref{sec:real_time}, we introduce the challenge of real-time voice authentication, how it is
performed, and why the speaker
models need to be trained to match the test statistics. In
Section \ref{sec:metric}, we use window-based EER to 
 evaluate the performance of the proposed AVA system.
In Section \ref{sec:ava_data}, we discuss the registration and
the data collection procedure. In Section \ref{sec:ava_ivector}, we evaluate the
i-vector technique for the AVA task. In Section \ref{sec:train}, we
introduce the architecture of the training and registration modules of the
AVA system and the algorithms that are applied to its
major components. In Section \ref{sec:test}, we discuss sequential
testing in the AVA system. In Section \ref{sec:expr},
we provide the evaluation results of the AVA system with
different configurations and algorithms.

\section{System Description and Technical Issues}
\label{sec:problem}

\subsection{Conventional Voice Authentication}
\label{sec:principle}
Use of a person's voice as a biometric indicator requires
processing of the signal to retain a salient representation of
the speaker-specific characteristics. Traditionally, these
may include the talker's source parameters (e.g., range
and dynamics of the pitch contour \cite{autocorrelation_pitch}, stress patterns) and
the tract parameters (e.g., the mean behavior of formant
frequencies, vocal tract length \cite{cepstrum_vt, lpc_vt}). 
Overall,
since these biometric parameters of the voice production
system represent a talker's intrinsic articulatory
characteristics, a substantial duration of the speech signal
is necessary, often in tens of minutes or even hours \cite{long_term_feature},
to support reliable estimation.


With advances in statistical modeling techniques, such
as the hidden Markov model (HMM), spectral features have become the
dominant choice to discriminate talker-specific voice characteristics
\cite{hmm_furui, hmm_sarkar}.
This has allowed a relative decrease in the duration of the speech
material required for training and testing, though it still remains
impractical for real-time monitoring applications. 
To address this problem, the traditional authentication approach is to use a likelihood ratio test
with MAP adapted universal background models (UBM) \cite{spkr_id_rose,
ubm_map} which
are built using Gaussian mixture models (GMMs). Adaptation techniques are
used to update the parameters of a pre-trained model using the new speech
signal. Further, discriminative training methods are applied to refine the speaker models with the goal of maximizing the speaker verification performance. In \cite{mve_speaker}, MVE training is proposed to jointly estimate the target and anti-target speaker models so that the expected number of verification errors (miss detection and false alarm) on enrollment and training set are minimized. Similarly, in \cite{liu1995study, angkititrakul2007discriminative}, minimum classification error (MCE) criterion \cite{mce} is used for speaker recognition and identification.  Based on these, the application of SVM in a speaker's GMM supervector space \cite{svm_1, svm_3} yields interesting results by performing a nonlinear mapping from the input space to an SVM extension space. 

More recently, factor analysis methods such as joint factor analysis
(JFA) \cite{jfa_2, jfa_3} and i-vectors \cite{ivector_1,
ivector_3} become the dominant approach for speaker verification. These approaches try to model the speaker and
channel variability by projecting speaker dependent GMM mean supervectors
onto a space of reduced dimensionality.
In recent years, deep vector (d-vector) \cite{variani2014deep} approach has achieved state-of-the-art
performance in NIST SREs, in which a deep neural network (DNN) is trained to classify speaker identities given their voice at the input. 
A d-vector is extracted per utterance by averaging the DNN hidden units to represent a registered speaker or a test utterance for subsequent speaker verification. 
Further, an end-to-end loss \cite{heigold2016end} and a triplet loss \cite{li2017deep} are introduced to learn more relevant embeddings to the speaker verification task. An attention mechanism is applied to dynamically summarize the DNN hidden units into speaker embeddings \cite{rahman2018attention}. To improve the noise robustness, DNN-based speaker embedding is further extended to x-vector in \cite{snyder2018x} by performing data augmentation. More recently, adversarial learning \cite{gan} with gradient reversal network \cite{ganin2015unsupervised} has been applied to domain adaptation \cite{sun2017unsupervised, meng2017unsupervised} and domain-invariant training \cite{shinohara2016adversarial, meng2018speaker, meng2018adversarial} of the DNN acoustic model \cite{hinton2012deep}. Similarly, it can effectively improve the robustness of the speaker embeddings by jointly optimizing the DNN speaker classifier and an auxiliary disriminative network to mini-maximize an adversarial objective \cite{wang2018unsupervised, meng2019asv}.

However, these methods are specifically designed and well suited only for
speaker verification tasks within the NIST 
SRE framework, in which long speech utterances are used as the material
for single individual tests, ranging in duration from 10s to a few
minutes depending on the specific task (e.g., see the NIST speaker
verification tasks in the years 2000-2010 \cite{nist_website}.). More
specifically, these techniques work well only for modeling the long-term
statistical characteristics of a speaker, which does not coincide
with the short-time testing condition required by the AVA task. In
Section \ref{sec:ava_ivector}, we show that the AVA system based on
i-vector achieves an excellent authentication performance when the
duration of the test window is long enough. But the performance degrades
rapidly as the test window duration decreases. In general, many i-vector
based systems exhibit sharp performance degradation
\cite{ivector_short_2011, ivector_short_2012, poddar2019quality, ivector_short_kheder}, when they are tested with
short duration (below 5s) utterances. This is understandable as the
covariance matrix of the i-vector is inversely proportional to the number
of speech frames per test utterance and the variance of the i-vector
estimate grows directly as the number of frames in the test utterance
decreases \cite{jfa_3}.


Recently, many approaches have been proposed for speaker verification with short-duration utterances. By borrowing the idea from speaker-adaptive training, the authors of \cite{soldi2014short} propose phone adaptive training (PAT) to learn a set of transforms that project features into a phoneme-normalized but speaker-discriminative space and use the normalized feature to improve speaker modeling given short-duration enrollment data. To alleviate the large estimation variation of i-vector due to short-duration utterances, uncertainty propagation is introduced to both the PLDA classifier \cite{stafylakis2013text} and the i-vector extraction \cite{kenny2013plda}. However, these methods only show their effectiveness on test utterances of about 3 seconds duration and can hardly meet the real-time requirement of AVA. To further overcome the mismatched prior distributions of the data used to train UBM and short-duration enrollment data, \cite{li2016improving} divides the speech signal into several subregions defined by speech unit and perform speaker modeling and verification within each subregion. A good improvement is achieved over GMM-UBM baseline for test utterances no longer than 2 seconds. However, these systems are rather complicated which entail large computations during testing and may lead to non-negligible delays in making real-time decisions of AVA.







\subsection{Challenges in Real-Time Voice Authentication}
\label{sec:real_time}

Most speech processing systems follow the convention of the
short-time analysis (STA) framework, in which segments of
signal, each being called a \emph{speech frame} with a duration
(denoted by $T_f$) of $20-40$ ms, are successively extracted for
analysis. 
The successive analysis is performed at a
predefined rate, called the frame rate denoted as $r_f$, a prevalent
choice of which is 100 per second. The frame rate can be
converted to frame shift, $\delta_f$, which is the reciprocal of
$r_f$.
 

The continuous monitoring mode of AVA dictates that it must be 
\emph{text-independent}, and it must perform \emph{real-time}
authentication sequentially, \emph{continuously} reporting the near-instantaneous
authentication results in preparation for possible breach
of prior authentication at any moment. The major challenge in designing such a system
is to effectively train talker-specific models, using as little enrollment speech as possible, for accurate, continuous and
instantaneous \emph{text-independent} speaker verification, with very
short test signals.

\begin{figure}[htpb!]
	\centering
	\includegraphics[width=0.7\columnwidth]{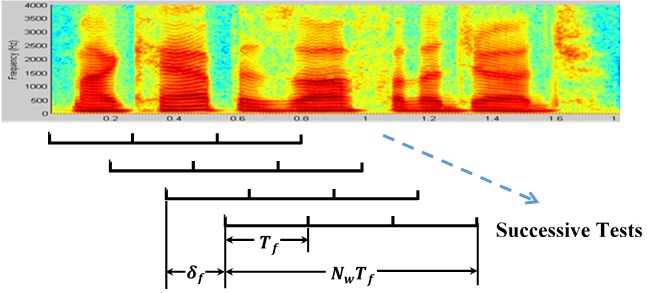}
	\caption{An illustration of the successive tests performed with
		data windows on a short-time spectrogram.} \label{fig:sliding_window}
\end{figure}

Since a talker change may happen abruptly, the authentication
decision cannot be based on a long memory of both the signal
representations and the prior decisions. 
However, it is well known in statistical analysis that more data means
better test results.  A trade-off is thus necessary in determining the
duration of data, which is subject to successive authentication tests.
This duration will involve multiple aforementioned frames as the typical
analysis frame length of $20-40$ ms is known to be far too short for
reliable hypothesis testing. We shall call the test segment a
\emph{``window''}, which is expressed in number of frames, $N_w$, and is
equivalent to $(N_w - 1)\delta_f + T_f$ of signal in time. As the system
slides the test ``window'' through the sequence of frame-based
representations and obtain the corresponding test scores, the reporting
interval then defines how often these scores need to be reported. In
other words, the temporal resolution for authentication test may not be
identical to that for reporting. Fig.~\ref{fig:sliding_window}
illustrates the concept of analysis frames and test windows.

To make accurate decisions, we need to model the speaker characteristics
within the specified short-time test windows. An unconventional speaker
modeling concept is in order here due to the aforementioned short-test
condition. In the usual hidden Markov modeling of speech for speaker
identification or verification, the model is implicitly assumed to be a
characterization of the \emph{general statistical behavior of the signal
source, without any regard to the test duration}, and the likelihood
calculation is a mere accumulation of the frame likelihoods over whatever
length of the given utterance might be. This means the models so trained
in the conventional manner, without a definitive duration of data, will
have an inherent mismatch condition in the captured statistical
knowledge, and may not lead to the most reliable likelihood test results.
To deal with this problem, we have to match the training and testing
condition by extracting short-time speech segments with a matching duration
from the training and enrollment speech feature sequence. The
speech segment within a sliding window at each time will serve as a
training token for HMM. In this case, the talker-specific HMM, which includes a pair of target and anti-target models, is able to model
the short-time characteristics of a speaker that is required by
the real-time testing process.

To meet the challenge of minimal
enrollment, we adapt a speaker-independent (SI) model with the MAP adaptation 
technique \cite{map} to make use of
the limited adaptation data and to obtain a decent estimation of the
model parameters with prior knowledge about the target model
distribution. In addition, the method of MVE training \cite{mve_speaker} is applied so that the total verification error is minimized. 
MVE in \cite{mve_speaker} must 
be adapted to the current operating setup of short test signals. It means 
that the notion of empirical error estimate in discriminative methods 
must now be based on an implicitly different test sampling scheme. It is no 
longer an utterance-based sampling and thus the inherent test statistics 
must be interpreted differently. We first address these design changes 
from the viewpoint of performance metric in Section \ref{sec:metric}. We also 
address the problem of data imbalance typical of MVE based systems
by pre-selecting a \emph{cohort set} consisting of the most confusing
impostor data. This balances the amount of target and impostor data
and expedites the time required for MVE training.
 

\subsection{Performance Metrics}
\label{sec:metric}
In traditional speaker verification (e.g., NIST SREs), the error counts are accumulated from the \emph{utterance-level} decisions: a ``miss'' occurs when a legitimate talker is denied for the entire test utterance and a ``false alarm'' occurs when an impostor is incorrectly accepted for the utterance. The EER is defined as the rate at which
the ``miss'' rate and the ``false alarm rate'' are equal \cite{nist_eer}.
This utterance-based error counting is
obviously not suitable for the AVA task because it bears the imperative
assumption that the entire test speech signal is uttered by one and only
one talker. It produces a single verification decision over the entire
utterance without considering the possible change of speaker identity
within the test signal. As noted above, AVA has to be prepared to detect
a change of talker at any moment, and a user authentication error may
occur at every test window slided over the signal continuously. 

For AVA, we evaluate \emph{window-based} EER (WEER) because each real-time decision about the user identity is made
on a test window anchored at that time instant. A window-based miss
detection error (WMDE) occurs if a ``reject'' decision is made while the authorized talker
is actually speaking within that window. A window-based false alarm error
(WFAE) occurs if an ``accept'' decision is made while an
impostor is speaking within that window. After all the window-based
testings are performed, the WMDE rate and the WFAE rate can be evaluated
against a chosen testing threshold. The WEER is reached when the
threshold is chosen to make the two error rates equal. Obviously,
calculation of the WMDE and WFAE rates is conditioned on the voice
activity detector; when there is no speech, no decision is to be
included. Note that WEER differs from the conventional utterance-based EER only in that the error counts are collected from window-level decisions instead of utterance-level ones. It becomes the traditional EER when each window of speech is treated as a separate test utterance.
 
With WEER as the performance metric, training of the models in AVA must
match the short-time testing condition, particularly when discriminative
modeling methods are used.  The purpose of discriminative model training
is to minimize the empirical error rate. For an AVA system, such an
empirical error rate is calculated from a combination of the WMDEs the
WFAEs (See Eq.~\eqref{eqn:average_cost}). All these authentication errors
are based on the window-based tokens. Therefore, the sample tokens for
training and enrollment must each correspond to a segment of speech signal within a test
window. This is one of the crucial differences in modeling for an AVA
system and for a conventional utterance based authentication system.

\section{AVA Database and Pre-Processing}
\label{sec:ava_data}



Since AVA is a different task from the conventional speaker verification directed by NIST SRE, we collect a new voice database, which we call the AVA
database, from 25 volunteers (14 females, 11 males) for performance evaluation. A Microsoft Surface
Pro tablet with a built-in microphone was used to record the data and the
sampling rate was set to 8000 samples/s. Each talker speaks at any
position relative to the device as he or she feels comfortable; we
consider this a natural use configuration of the device.  The data
collected from each person consists of four parts: the rainbow
passage~\cite{rainbow}, a user-chosen pass-phrase, 20 randomly selected
sentences from the phonetically balanced Harvard sentences~\cite{harvard}
(5.5~s on average) and 30 digit pairs (each digit is randomly selected
from 0 to 9). The speaker repeats the same pass-phrase 8 times. In total,
the recording amounts to 2.5 hours of voice signal from all talkers.

For each speaker, we choose the enrollment data from the
Rainbow passage, the pass-phrases and digits while the testing data is
chosen from the Harvard sentences. The enrollment and test data sets do
not overlap. 
The duration of each test set is configured to provide at
least 1000 decisions per speaker in any given configuration. In all the
experiments of this paper, the audio signal is converted to the
conventional 39-dimension MFCC features with frame duration $T_f=25$~ms
and $\delta_f=10$~ms. For the AVA task, the enrollment or test window moves
forward 10~ms each time. Successive tests are performed with each shift
over a segment of the specified durations. The durations of the enrollment
and test windows are equal. The cepstral mean of speech frames within
each enrollment and test window is subtracted to minimize the channel
variability.


\section{AVA with I-Vector}
\label{sec:ava_ivector}

I-vector analysis, a new front end factor analysis technique,
is the predominant tool for conventional speaker verification.
In this section, we investigate if this widely applied technique
can achieve satisfactory performance for AVA.

The i-vector is a projection of a speech utterance onto a
low-dimensional total variability space that models both the
speaker and the channel variability. More specifically, it is
assumed that there exists a linear dependence between the
speaker adapted (SA) GMM supervectors $\boldsymbol{\mu}$ and the SI 
GMM supervector $\boldsymbol{m}$ \cite{ivector_1}.

\begin{align} \boldsymbol{\mu}=\boldsymbol{m}+\boldsymbol{U}\boldsymbol{w}
	\label{eqn:ivector_define} \end{align}
where $\boldsymbol{U}$ is a low rank factor loading matrix representing
the primary direction of variability, and $\boldsymbol{w}$ is a random
vector of total factors having a standard normal distribution
$\mathcal{N}(\boldsymbol{0}; \boldsymbol{I})$. The i-vector is an MAP estimate of $\boldsymbol{w}$.

%

We first apply i-vector to the conventional speaker verification task
under the assumption that each test utterance is from only one speaker.
We train a GMM UBM with all the enrollment
data in the AVA database. With the EM algorithm, an SI
factor loading matrix $\boldsymbol{U}_{SI}$ is trained on the statistics collected
from the UBM. An i-vector is then extracted for each speaker using his or
her enrollment data and $\boldsymbol{U}_{SI}$. During testing, an i-vector is extracted
from the each test utterance using $\boldsymbol{U}_{SI}$. The i-vector dimension is fixed at 400. A cosine distance between
the i-vector of each test utterance and that of the hypothesized speaker
is used as the decision score. The EER is computed with all the
utterance-level decision scores. In AVA database, the i-vector achieves
0.00\% EER for the utterance-based speaker verification task under all
UBM configurations.

We then apply i-vector for the AVA task. We adopt the same training method
as in the traditional speaker verification except that the training and enrollment tokens are generated by a
sliding window with a prescribed duration. During testing, a test window
of the same duration is slided over the test utterance at the rate of 100
per second and an i-vector is extracted from the speech signal within
each test window using $\boldsymbol{U}_{SI}$.  The cosine distance between the
i-vector of each test window and that of the hypothesized speaker is used
as the decision score.


The AVA dataset described in Section \ref{sec:ava_data} is used for the
performance evaluation. We fix the duration of enrollment data at an
average of 240~s per speaker and randomly select two Harvard sentences
for use as the testing data for each speaker. For AVA task, the window
duration ranges from 1.01~s to 3.01~s. We show the WEER results with
respect to the test window duration and the number of mixtures in the UBM
in Table \ref{table:ivector_ava}.

\begin{table}[htbp!]
\centering
\caption{WEER (\%) of AVA using i-vector on AVA database with
different test window durations and UBM configurations. The
enrollment data is 240~s long on average for each speaker.}
\newcolumntype{L}[1]{>{\raggedright\let\newline\\\arraybackslash\hspace{0pt}}m{#1}}
\newcolumntype{C}[1]{>{\centering\let\newline\\\arraybackslash\hspace{0pt}}m{#1}}
\newcolumntype{R}[1]{>{\raggedleft\let\newline\\\arraybackslash\hspace{0pt}}m{#1}}
\resizebox{0.85\columnwidth}{!}{
	\begin{tabular}[c]{C{2.5cm}||C{1.5cm}|C{1.5cm}|C{1.5cm}|C{1.5cm}|C{1.5cm}}
	\hline
	\hline
	\multirow{2}{*}{\begin{tabular}{@{}c@{}}Number \\ of Mixtures
	\end{tabular}} & \multicolumn{5}{c}{Test Window Duration (s)} \\
	\hhline{~-----}
	& 1.01 & 1.51 & 2.01 & 2.51 & 3.01 \\
	\hline
        \hline
        64 & 14.82 & 7.97 & 4.31 & 1.96 & 0.87 \\ 
	\hline
	128 & 13.72 & 7.24 & 3.72 & 1.56 & 0.58 \\
	\hline
	256 & 13.89 & 7.29 & \textbf{3.69} & \textbf{1.43} &
	\textbf{0.35} \\
	\hline
	512 & \textbf{12.91} & \textbf{6.92} & 3.79  & 1.44 & 0.52 \\
	\hline
	1024 & 14.54 & 8.02 & 3.99 & 1.62 & 0.64 \\
	\hline
	\hline
        \end{tabular}
}
	\label{table:ivector_ava}
\end{table}
For each UBM configuration, the i-vector based AVA system achieves
\textless 1\% WEER when the duration of the test window is above 3~s. The
performance degrades drastically as the test window duration falls below
2~s. When the test window is 1~s, the WEER rises to 12.91\%. This
performance trend is consistent with what have been reported in the
literature and we conclude that it is not suitable for the AVA task where
accurate decisions about speaker identity need to be made
instantaneously.

\section{AVA Training and Registration}
\label{sec:train}
The AVA system consists of three parts: a training module, a registration module and an authentication module. In
this section, we introduce the major components of the training and registration module which train and adapt the models to the enrollment data of each speaker.

\begin{figure}[htpb!]
	\centering
	\includegraphics[width=0.8\columnwidth,height=0.80\columnwidth]{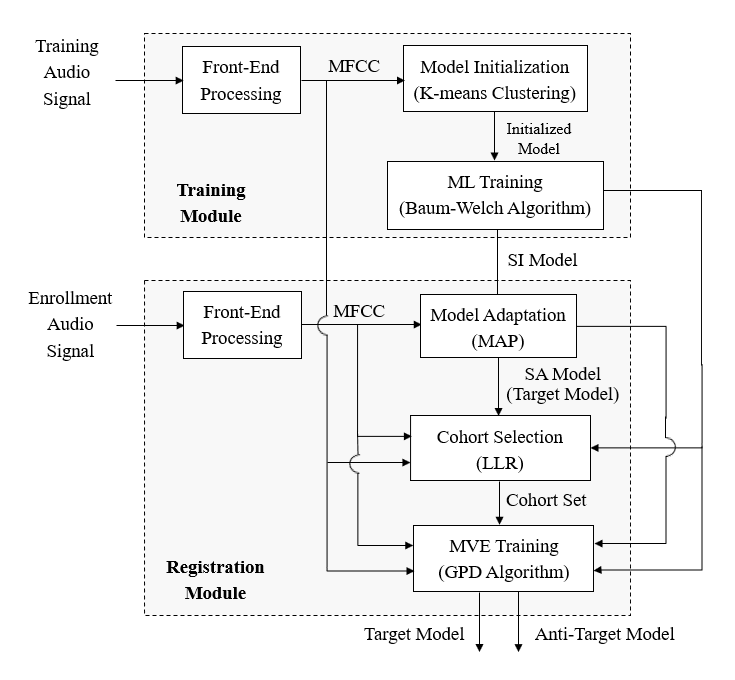}
	\caption{The components of the AVA training and registration stages.}
	\label{fig:ava_train}
\end{figure}

Fig.~\ref{fig:ava_train} shows the training and registration stages of the AVA system.
First, in the training stage, a SI ergodic HMM is trained on a sufficient pool of data from a general collection of speakers in the training set. 
The speech signal is converted to 
mel-frequency cepstral coefficients (MFCCs)
\cite{mfcc} through the front-end processing component. The parameters
of the SI HMM are initialized with the K-means
clustering algorithm. The final SI HMM is obtained by
applying the \emph{Baum-Welch} re-estimation algorithm in the maximum
likelihood (ML) training component. Then, in the registration stage, the model adaptation component 
adapts the SI model parameters to the voice of the target 
speaker upon receipt of the new registration data and generates the SA model based on the MAP adaptation technique
\cite{map}. Then for the target data, an equivalent
and most confusing set of data is selected from the impostor set by the
cohort selection component for MVE training. Finally, the MVE training
component generates the MVE trained target and anti-target model by directly
minimizing a combination of the WMDEs and WFAEs. We elaborate
the algorithms and procedure in Section \ref{subsec:train_si}, \ref{subsec:train_map}, 
\ref{subsec:train_mve} and \ref{subsec:train_cohort}.


\subsection{Speaker-Independent (SI) Model Training}
\label{subsec:train_si}
In the training stage, an SI ergodic HMM, also called the UBM, for use as the seed model for later adaptation is trained. The
parameters of this HMM are estimated by the \emph{Baum-Welch}
re-estimation algorithm after having been initialized 
with the K-means clustering algorithm.

Let $\gamma_j (t)$ denote the occupation probability of being in state $j$ of the
ergodic HMM at time $t$ which can be calculated efficiently using the
\emph{Forward-Backward} algorithm \cite{hmm}. The above-mentioned
short-time requirement in sequential training implies that $\gamma_j(t)$
be accumulated differently from the conventional utterance-based training
approach. This is because each test in AVA involves a voice segment
within a test window of duration, $N_w \delta_f$, and this condition
should be matched during training. Therefore, we modify the accumulation
of $\gamma_j(t)$ as follows.

Let us denote an entire training utterance by
$X=\{\boldsymbol{x}_1,\ldots, \boldsymbol{x}_T\}$,
where $T$ is the number of frames within the training utterance.
$X_{t_a,t_b}=\{\boldsymbol{x}_{t_a},\ldots,\boldsymbol{x}_{t_b}\}$ is the speech segment
extracted from $X$, where $t_a$ and $t_b$ are the start and end times,
respectively, and $1\le t_a \le t_b \le T$. Each $X_{t_a,t_b}$ is used
as a training token for \emph{Baum-Welch} re-estimation as is
elaborated above. Assume that the number of frames within each window
is $N_w$ and $s_t$ is the state that frame $\boldsymbol{x}_t$ is aligned with at
time $t$. 
In short-time sequential training, the occupation probability $\gamma^{short}_j(t)$ of being in state $j$ at time $t$ becomes
\begin{align}
	\gamma^{short}_j(t)&=\frac{1}{N_w}\sum_{\tau=1}^{N_w} P(s_t=j|X_{t+\tau-N_w+1,t+\tau}), \\
	&=\frac{1}{N_w}\sum_{\tau=1}^{N_w} \frac{P(X_{t+\tau-N_w+1,t+\tau}|s_t=j)P(s_t=j)}{P(X_{t+\tau-N_w+1,t+\tau})}.
	\label{eqn:occ_win}
\end{align}
In Eq.~\eqref{eqn:occ_win}, $\gamma^{short}_j(t)$ is calculated through the average
likelihood of all the speech segments of window duration $N_w\delta_f$ which
include the frame $\boldsymbol{x}_t$.

For the conventional utterance-based training, the state occupation is
\begin{align}
	\gamma^{conv}_j(t)&=P(s_t=j|X_{1,T}), \\
	&=\frac{P(X_{1,T}|s_t=j)P(s_t=j)}{P(X_{1,T})}.
	\label{eqn:occ_utt}
\end{align}
In Eq.~\eqref{eqn:occ_utt} $\gamma^{conv}_j(t)$ is computed through the
likelihood of the entire utterance $X$, which is much longer than the
window duration. Each term of the summation in Eq~\eqref{eqn:occ_win} falls
back to the conventional utterance-based state occupation probability in Eq~\eqref{eqn:occ_utt} when $N_w=T$ and $\tau=T-t$ because the window covers the duration of the entire utterance.
In other
words, $\gamma^{short}_j(t)$ is affected by the statistics of the
window-duration speech segment which contains frame $\boldsymbol{x}_t$ while
$\gamma^{conv}_j(t)$ is affected by the entire training utterance even when
only a small portion of the utterance is correlated with $\boldsymbol{x}_t$
statistics.

After obtaining the occupation probability, we update the HMM
parameters with the average of the window-wise sufficient statistics
weighted by $\gamma^{short}(t)$ in a standard way.

\subsection{Model Adaptation}
\label{subsec:train_map}
When a registration procedure is initiated, the 
SI model is assumed to have been well-trained as described in Section \ref{subsec:train_si}.
As the first step of registration, the model adaptation component in Fig.~\ref{fig:ava_train}
adapts the SI model to the new registration data of the
authorized target speaker 
using MAP estimation. 

Assuming speech
segment $X=\{\boldsymbol{x}_1,\ldots,\boldsymbol{x}_T\}$ within a sliding window from a registered user to be a training token for MAP adaptation,
the likelihood of $X$ given HMM with $J$
states and parameter $\lambda=\{{\pi_j},a_{ij},\theta_j\}_{i,j=1}^J$ is
\begin{equation}
	p(X|\lambda)=\sum_{\boldsymbol{s}}\pi_{s_0}\prod_{t=1}^Ta_{s_{t-1}s_{t}}\sum_{m=1}^Kw_{s_tm}\mathcal{N}(\boldsymbol{x}_t|\boldsymbol{\mu}_{s_tm},\boldsymbol{\Sigma}_{s_tm})
	\label{eqn:likelihood_hmm}
\end{equation}
where $s=\{s_1,\ldots,s_T\}$ is the unobserved state sequence, $\pi_j$
is the initial probability of state $j$, $a_{ij}$ is the transition
probability from state $i$ to state $j$, 
$\theta_j=\{w_{jk},\boldsymbol{\mu}_{jk},\boldsymbol{\Sigma}_{jk}\}$, $k=1,\ldots,K$, where
$w_{jk}$, $\boldsymbol{\mu}_{jk}$, $\boldsymbol{\Sigma}_{jk}$ are the weight, mean vector and
covariance matrix, respectively, for the $k$ th component of the
Gaussian mixture which is the probability output of state $j$.
The MAP estimate $\theta_{MAP}$ is aimed at maximizing the posterior
probability denoted as $f(\lambda|X)$, i.e.,
\begin{equation}
	\theta_{MAP}=\arg\max_\lambda f(\lambda|X)=\arg\max_\lambda p(X|\lambda)g(\lambda)
	\label{eqn:map}
\end{equation}
where $g(\lambda)$ is the prior distribution of $\lambda$.

The MAP estimate is obtained as follows. The probability of being in
state $j$ at time $t$ with the $k$th mixture component accounting for
$\boldsymbol{x}_t$ is
\begin{equation}
	\gamma_{j,k}(t)=\gamma_j(t)\frac{w_{jk}\mathcal{N}(\boldsymbol{x}_t|\boldsymbol{\mu}_{jk},\boldsymbol{\Sigma}_{jk})}{\sum_{m=1}^Kw_{jm}\mathcal{N}(\boldsymbol{x}_t|\boldsymbol{\mu}_{jm},\boldsymbol{\Sigma}_{jm})}
	\label{eqn:gamma}
\end{equation}

For mixture $k$ in state $j$, the occupation likelihood and the 1st
and 2nd moment of the observed adaptation data can be estimated by
\begin{align}
	&n_{jk}=\sum_{t=1}^{T}\gamma_{j,k}(t), \quad E[{\boldsymbol{x}_t}]=\frac{1}{n_{jk}}\sum_{t=1}^T\gamma_{j,k}(t)\boldsymbol{x}_t \\
	&E[{\boldsymbol{x}_t}{\boldsymbol{x}_t}^\top]=\frac{1}{n_{jk}}\sum_{t=1}^T\gamma_{j,k}(t)\boldsymbol{x}_t\boldsymbol{x}_t^\top 
	\label{eqn:mom}
\end{align}
Thus, the MAP update formula for mixture $k$ in state $j$ of an HMM is 
\begin{align}
	&\hat{w}_{jk}=\alpha_{jk}^w\frac{n_{jk}}{T}+(1-\alpha_{jk}^w)\bar{w}_{jk} \\
	&\hat{\boldsymbol{\mu}}_{jk}=\alpha_{jk}^m
	E[\boldsymbol{x}_t]+(1-\alpha_{jk}^m)\bar{\boldsymbol{\mu}}_{jk} \\
	&\hat{\boldsymbol{\Sigma}}_{jk}=\alpha_{jk}^v
	E[\boldsymbol{x}_t\boldsymbol{x}_t^\top]+(1-\alpha_{jk}^v)(\bar{\boldsymbol{\Sigma}}_{jk}+\bar{\boldsymbol{\mu}}_{jk}^2)-\hat{\boldsymbol{\mu}}_{jk}^2
	\label{eqn:update}
\end{align}
where $\{\bar{w}_{jk},\bar{\boldsymbol{\mu}}_{jk},\bar{\boldsymbol{\Sigma}}_{jk}\}$, $k=1,\ldots,K$, $j=1,\ldots,J$ are the mixture parameters of the SI HMM. 
The adaption coefficient $\alpha_{jk}^\rho,\rho \in\{w,m,v\}$ is defined
for each mixture component in each state as
$\alpha_{jk}^{\rho}=n_{jk}/(n_{jk}+\eta^\rho)$
where $\eta^\rho$ accounts for the weight of prior knowledge for $\rho$.

The MAP adaptation is performed during sequential training on
each training token $X$. As is explained in
Section~\ref{subsec:train_si}, the MAP adapted model also characterizes the
short-time statistics of the registration speech data since the statistic
$\gamma_j(t)$ in Eq.~\eqref{eqn:gamma} is accumulated through the
likelihoods of the adaptation speech segments which have the same
duration as the test window.
\subsection{MVE Training}
\label{subsec:train_mve}
As registration stage, the MVE training is performed after the speaker model
adaptation. The SA HMM and the
SI HMM serve as the initial target and initial
anti-target model, respectively, for the MVE training. In the MVE
training component in Fig.~\ref{fig:ava_train}, all parameters are
optimized with the enrollment and training data, according to the
criterion to minimize the total
number of authentication errors (which is the total number of WMDEs and
WFAEs) on the corpus.


 Let us define the enrollment data of the target speaker as the target set $D_0$, and define the training data excluding the speech of the target speaker as the impostor set $D_1$. For a window-duration MVE training token $X_n$ from either $D_0$ or $D_1$, $g(X_n|\lambda_0)$ and
$g(X_n|\lambda_1)$ denote the log-likelihoods of $X_n$ given the target HMM with parameters $\lambda_0$ and the anti-target HMM with
parameters $\lambda_1$, respectively. The log-likelihoods are
calculated by aligning $X_n$ against the states of the target and the
anti-target models using the \emph{Viterbi} algorithm and are
normalized with respect to the total number of frames $T$ within the
utterance $X_n$.  As the training tokens are generated by sliding a
window of size $N_w$ frames every $\delta_f$ duration on the training
utterance, the likelihood can be calculated more efficiently by
modifying the \emph{Viterbi} algorithm. Instead of resetting the
trellis and initializing it anew each time we evaluate the
log-likelihood of a new window of voice segment within the same
utterance, we reset the trellis only at the beginning of a training
utterance and let the trellis grow until the end of the utterance. The
log-likelihood of an incoming window of voice segment is accumulated
directly from the part of the fully grown trellis which starts from
the very beginning of the utterance. This new implementation is
equivalent to performing a partial traceback of the trellis structure
within each sliding window so that the consistency is maintained in
training and testing based on short window of data. The partial
traceback also speeds up the MVE training procedure by a factor of
$N_w$.



To count the verification errors based on the log-likelihood of the 
tokens, the \emph{misverification} measure is further defined for
each class
\begin{align}
	d_0(X_n|\lambda_0,\lambda_1)&=-g(X_n|\lambda_0)+g(X_n|\lambda_1), \quad \text{if } X_n \in D_0, \label{eqn:misverification_1} \\		
	d_1(X_n|\lambda_0,\lambda_1)&=g(X_n|\lambda_0)-g(X_n|\lambda_1), \quad \text{if } X_n \in D_1. \label{eqn:misverification_2}
\end{align}
The two types of verification errors, WMDE and WFAE, can be approximated as $l_0$ and $l_1$, respectively, by embedding the two misverification measures into smooth loss functions below
\begin{align}
	l_0(X_n|\lambda_0,\lambda_1)&=\frac{A_0}{1+\exp\left[-d_0(X_n|\lambda_0,\lambda_1)\right]}, \quad \text{if } X_n \in D_0,  \label{eqn:loss_function_1}\\		
	l_1(X_n|\lambda_0,\lambda_1)&=\frac{A_1}{1+\exp\left[-d_1(X_n|\lambda_0,\lambda_1)\right]}, \quad \text{if } X_n \in D_0, \label{eqn:loss_function_2}
\end{align}
where the weights $A_0$ and
$A_1$ emphasize the respective error
types. 

Finally, we obtain the MVE loss as an
approximation of the total number of verification errors on the training and enrollment corpus as follows.
\begin{equation}
	L(\lambda_0,\lambda_1)=\sum_{X_n \in D_0} l_0(X_n |\lambda_0,\lambda_1)+\sum_{X_n \in D_1} l_1(X_n |\lambda_0,\lambda_1)		
	\label{eqn:average_cost}
\end{equation}
In Eq.~\eqref{eqn:average_cost}, the total number of
verification errors are expressed as a
continuous and differentiable function of the model parameters, and
hence, can be minimized with respect to all parameters by using the
generalized probabilistic descent (GPD) algorithm \cite{gpd}. 

In the short-time sequential training framework, each $X_n$
is a speech segment sequentially extracted from the
training utterance via a sliding window. The likelihoods of the speech segments are
computed with the same duration as the test window and then utilized
to calculate the gradient and the steps size of GPD update at each
iteration. Therefore, a speaker model accurately matched to the
testing condition is estimated through short-time sequential
training. 


\subsection{Cohort Selection}
\label{subsec:train_cohort}
From Eqs.~\eqref{eqn:misverification_1},
\eqref{eqn:misverification_2},~\eqref{eqn:loss_function_1}~and~\eqref{eqn:loss_function_2}, 
we notice that an MVE training frame from the target speaker updates the
model such that the WMDE decreases and a training token from the
impostor speaker updates the model such that the WFAE 
decreases. The impostor speaker can be any speaker other than the target
speaker and the number of data tokens from the impostors in the database will 
obviously be greater than the number of tokens of the target
speaker. Since we choose WEER as the performance indicator, we need a
balanced data set of target and impostor
speech. 
Therefore, to maintain a balance in the amount of target and impostor
data without sacrificing the discriminability between the two, we pick
from the impostor set $D_1$ a most confusing set of data for use in MVE
training for each target speaker. This selected set is called the
\emph{cohort set}.


In the cohort selection component of Fig.~\ref{fig:ava_train}, a
screening test is run with the MAP adapted target model and the
SI model to select possible cohort impostor set for MVE
training. A log-likelihood ratio (LLR) is computed for each
window-duration segment of the impostor data $D_1$ using
Eq.~\eqref{eqn:likelihood_hmm}, where the log-likelihoods are
calculated efficiently through the modified \emph{Viterbi} algorithm
described in Section~\ref{subsec:train_mve}. We further rank the
segments by their LLRs in descending order and pick the top $r$ speech
segments as the cohort set for subsequent MVE training, where $r$ is
the number of segments in the target dataset. This does not affect the
overall performance as the speech segments with lower LLRs naturally
contribute less to the gradient in GPD optimization (see
Eqs.~\eqref{eqn:misverification_1},~\eqref{eqn:misverification_2},~\eqref{eqn:loss_function_1},
\eqref{eqn:loss_function_2}~and~\eqref{eqn:average_cost}).
With cohort selection, $D_1$ in Eq.~\eqref{eqn:average_cost} becomes the cohort set and the following MVE training is performed in the same way as described in Section \ref{subsec:train_mve}. Cohort selection proves useful in reducing
the time needed to train the speaker models with MVE as it reduces the
data that the MVE algorithm needs to process.

\section{Short-Time Sequential Testing}
\label{sec:test}

\begin{figure}[htpb!]
	\centering
	\includegraphics[width=0.7\columnwidth, height=0.4\columnwidth]{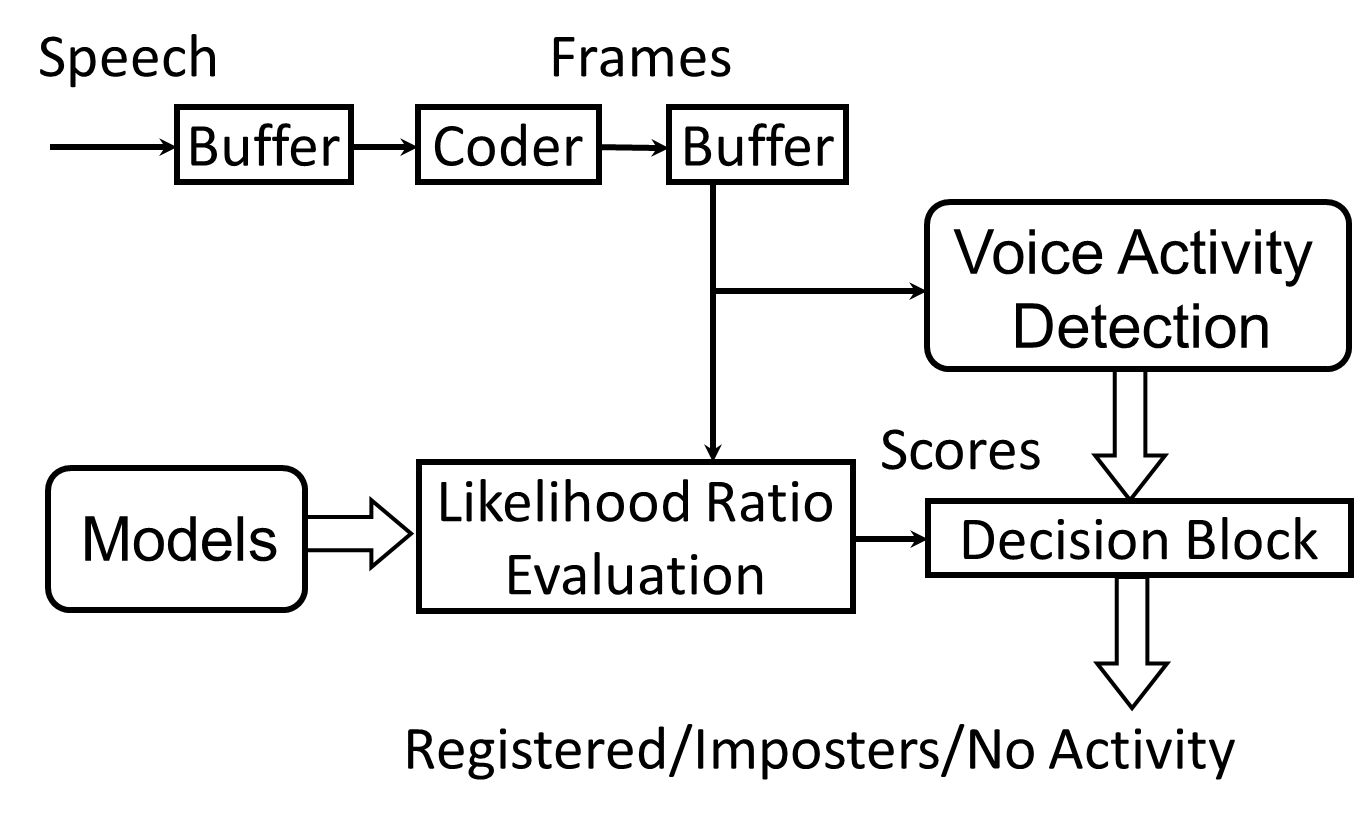}
	\caption{Diagram of the AVA authentication module.}
	\label{fig:ava_monitor}
\end{figure}

In the authentication stage of AVA, the system performs sequential testing
(note: the sequential testing here is to be differentiated from the
Wald's sequential test \cite{wald}) and makes decisions in
real-time. The
testing needs both the target and anti-target models for each
registered speaker that are obtained after MVE training. During
operation, the sequential testing procedure continuously takes a
sliding window of speech frames, accumulates the log-likelihood with
respect to both target and anti-model for the speaker, and then
reports the LLR confidence scores periodically to the
system. Fig.~\ref{fig:ava_monitor} shows the block diagram for short-time
sequential testing.
The LLR scores are calculated using the following equations
\begin{equation}
	\Gamma(X)=\log p(X|\lambda_0)-\log p(X|\lambda_1)
	\label{eqn:llr}
\end{equation}
where $X=\{\boldsymbol{x}_1,\ldots,\boldsymbol{x}_T\}$ is a window of voice frames. $\lambda_0$
and $\lambda_1$ are the parameters of the target and anti-target
models defined in \label{eqn:np_log}, respectively. The likelihoods $p(X|\lambda_0)$ and
$p(X|\lambda_1)$ are computed using Eq. (\ref{eqn:likelihood_hmm})
through the modified \emph{Viterbi} algorithm described in Section
\ref{subsec:train_mve}.

As discussed, a speech signal inevitably contains silence
gaps. These silence gaps do not contain any voice biometric
information and need to be excluded from testing. We use a
voice activity detector (VAD) to modulate the WEER results by ignoring
the test scores from silent frames. We use the VAD algorithm suggested
in the European Telecommunications Standards Institute (ETSI)
Distributed Speech Recognition front-end \cite{etsi}. The VAD makes a
binary voice/silence decision for every frame. Each VAD decision 
is made based on the average log mel energy of its 80 neighboring frames. We ignore the speaker
authentication decision for a given testing window if the
corresponding anchor frame (the frame at the middle of window) is
silent according to the VAD.

\section{Experiments}
\label{sec:expr}
For the performance evaluation in Sections \ref{subsec:expr_win_dur},
\ref{subsec:expr_model_config}, and \ref{subsec:expr_data_dur}, we use
the enrollment and test data in AVA dataset as is described in Section
\ref{sec:ava_data}. 



\subsection{Performance with test window duration}
\label{subsec:expr_win_dur}
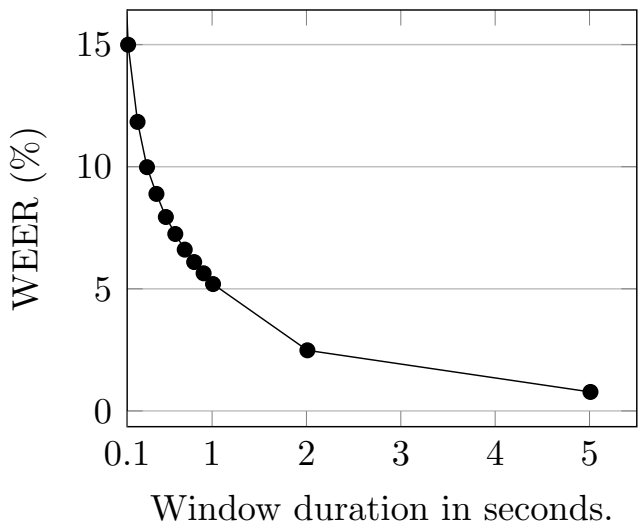
\begin{figure}[htpb] \centering
  \resizebox{0.65\columnwidth}{!}
  {
    \begin{tikzpicture}
      \begin{axis}[ 
        xlabel={\footnotesize Window duration in seconds.}, 
        ylabel={\footnotesize WEER (\%)},
        width=\columnwidth,
        small,
        xmin=0.1,
        extra x ticks={0.1},
        extra x tick labels={0.1},
        ymajorgrids=true
        ] 
        \addplot[black,mark=*] table {EERWinDur.dat};
      \end{axis}
    \end{tikzpicture} 
  }
  \caption{The WEER as a function of the duration of the
    decision window. The WEER values are based on the MAP adapted model
    as the target model and the SI model as the
    anti-target model with 1 state and 128 mixtures.}

	\label{fig:win_dur}
\end{figure}

Our first investigation focused on the trade-offs between the duration of 
the test window and the authentication performance. The duration of test 
window directly affects the system delay and the real-time requirement. 
We fix the duration of enrollment data at
an average of 240~s per speaker, but vary the duration of the test
speech segment from $N_w\delta_f=0.1$~s to $5.01$~s corresponding to
$N_w=1,\ldots, 501$
frames. We select part of the Harvard sentence set for use as the testing 
data for each speaker. Two Harvard sentences are randomly selected for each speaker
for the window durations from 0.1~s to 1.01~s, while 4
Harvard sentences are selected for 2~s and 5~s testing windows.   
Fig. \ref{fig:win_dur} shows the
baseline WEER for each window duration. We note that the WEER based on just
0.1~s of test data is quite poor at approximately 24\%, but it improves
as we increase the duration of the data for each decision
epoch.

As expected, the WEER performance monotonically decreases with the duration 
of the test data. Furthermore, a knee point can be observed at around 1~s, 
which can serve as a designing parameter to meet the real time requirement. 
We also note that the WEER with 5~s of sequential test data is 0.78\%.
This is quite low and in need of careful considerations as it may incorrectly 
imply that the system would perform flawlessly if the test window is sufficiently 
large. In our evaluation, we perform a test evaluation on each successive window 
of data which comes from the same utterance. For example, a 10~s long utterance 
will produce nearly 1000 test decisions; in the conventional utterance-based 
evaluation, it would have been just one test decision. This gives rise to the 
issue of statistical significance in the error probability estimate for the 
talker verification performance. It is fair to note that in the conventional 
utterance-based evaluation, the evaluation sample size tends to be rather 
limited, which weakens the statistical significance of the test result, while 
in the new window-based evaluation, the test sample size makes the error probability 
estimate more statistically trustworthy but it contains a sampling bias as the 
tests are performed on successive data windows that are not independent. This 
contrastive consideration, while interesting, does not affect the determination 
of the trade-off here. We thus choose 1~s as the nominal duration of the test 
data window, used in subsequent evaluations.


\subsection{Performance as a function of model configuration}
\label{subsec:expr_model_config}

We fix the duration of the enrollment data at an average of 240 s per
speaker and randomly select 2 or 4 Harvard sentences for each speaker as
the test set when the test window duration is 1.01 s or 2.01 s. With 25
speakers in total, more than 25,000 or 35,000 trials are generated from the 50
Harvard sentences or 100 Harvard sentences by sliding the test window.
In Table~\ref{table:w1_t240}, we provide the performance evaluations
for the average WEER (after VAD modulation) with a 1.01 second decision
window duration over the various model configurations and two
algorithms, MAP and MVE (the number in bold means the best performance
in the column). We notice that the average WEER for MAP adapted models
is 4.10\% while MVE training decreases the absolute WEER to 3.00\%, on
average. Depending on the complexity of the model and the algorithm,
the WEER ranges between 2.6-4.5\%. The models with one state
represented by 1024 Gaussian mixtures achieves the best
performance. Table~\ref{table:w2_t240} shows the evaluation results
with VAD for different configurations with the decision window
durations set at 2.01 seconds. 
\begin{table}[htbp!]
\centering
\caption{WEER performance under different HMM model configurations after VAD. The decision window duration is 1.01 seconds. The enrollment utterance is 240 seconds long on average for each speaker (full enrollment data).} 

\newcolumntype{L}[1]{>{\raggedright\let\newline\\\arraybackslash\hspace{0pt}}m{#1}}
\newcolumntype{C}[1]{>{\centering\let\newline\\\arraybackslash\hspace{0pt}}m{#1}}
\newcolumntype{R}[1]{>{\raggedleft\let\newline\\\arraybackslash\hspace{0pt}}m{#1}}
\resizebox{0.9\columnwidth}{!}{
	\begin{tabular}[c]{C{2.5cm}|C{2.5cm}|C{3.5cm}|C{3.5cm}}
	\hline
	\hline
	 Number of States & Number of Mixtures & MAP WEER (\%) & MVE WEER (\%) \\
	\hline
	\hline
	 1 & 128 & 4.51 & 3.17 \\
	 8 & 16 & 4.50 & 3.21 \\
	 \hline
	 1 & 256 & 4.13 & 2.89 \\
	 16 & 16 & 4.14 & 2.96 \\
	 \hline
	 1 & 512 & 3.76 & 2.85 \\
	 32 & 16 & 3.97 & 3.04 \\
	 \hline
	 1 & 1024 & \textbf{3.56} & \textbf{2.66} \\
	 32 & 32 & 4.22 & 3.22 \\
	 \hline
	 \hline
	 \multicolumn{2}{C{5cm}|}{Average} & 4.10 & 3.00 \\
	\hline
	\hline
\end{tabular}
}
\label{table:w1_t240}
\end{table}

\begin{table}[htbp!]
\centering
\caption{WEER performance under different HMM model configurations after VAD. The decision window duration is 2.01 seconds. The enrollment utterance is 240 seconds long on average for each speaker (full enrollment data).}
\newcolumntype{L}[1]{>{\raggedright\let\newline\\\arraybackslash\hspace{0pt}}m{#1}}
\newcolumntype{C}[1]{>{\centering\let\newline\\\arraybackslash\hspace{0pt}}m{#1}}
\newcolumntype{R}[1]{>{\raggedleft\let\newline\\\arraybackslash\hspace{0pt}}m{#1}}
\resizebox{0.9\columnwidth}{!}{
	\begin{tabular}[c]{C{2.5cm}|C{2.5cm}|C{3.5cm}|C{3.5cm}}
	\hline
	\hline
	 Number of States & Number of Mixtures & MAP WEER (\%) & MVE WEER (\%) \\
	\hline
	\hline
	1 & 128 & 2.10 & 1.55 \\
	8 & 16 & 2.15 & 1.45 \\
	\hline
	1 & 256 & 2.06 & 1.25 \\
	16 & 16 & 2.04 & 1.20 \\
	\hline
	1 & 512 & 2.07 & 1.29 \\
	32 & 16 & \textbf{1.97} & \textbf{1.12} \\
	\hline
	1 & 1024 & 2.31 & 1.87 \\
	32 & 32 & 2.26 & 1.39 \\
	\hline
	\hline
	\multicolumn{2}{C{5cm}|}{Average} & 2.12 & 1.39 \\
	\hline
	\hline
\end{tabular}
}
\label{table:w2_t240}
\end{table}
\subsection{Performance with enrollment data duration}
\label{subsec:expr_data_dur}
It is desirable to use as little enrollment data as possible while
maintaining a similar performance as in Table~\ref{table:w1_t240}. In
the following, we evaluate our system on the minimum amount of
enrollment voice data necessary to achieve an acceptable performance
which bears directly on the time it takes for a talker to register for
AVA for the first time.

In Table~\ref{table:w1_t105} and~\ref{table:w1_t180}, we use 180 and 105
seconds of enrollment voice data, respectively. Two Harvard sentences from
each speaker are selected to form the test set. With 25 speakers in
total, more than 25,000 trials are generated from the 50 Harvard
sentences by sliding a test window with a duration of 1.01 s. When we compare
the results in Table~\ref{table:w1_t180} with the results in
Table~\ref{table:w1_t240}, where the average enrollment data duration is
240 seconds, we notice that using 25\% less enrollment data reduces the MVE
performance to an average WEER of 4.64\% from 3.00\%. Similarly, in
Table~\ref{table:w1_t105} we are using 56\% less enrollment voice data than
in Table~\ref{table:w1_t240} which reduces the average WEER performance
with MVE to 6.31\%. In the case of reduced enrollment voice data, the
degradations may still be acceptable as the WEER stays in the vicinity of
5-6\%.

\begin{table}[htbp!]
\centering
\caption{WEER performance under different HMM model configurations after VAD. The decision window duration is 1.01 seconds. The \textbf{enrollment utterance is 180 seconds long} for each speaker.}
\newcolumntype{L}[1]{>{\raggedright\let\newline\\\arraybackslash\hspace{0pt}}m{#1}}
\newcolumntype{C}[1]{>{\centering\let\newline\\\arraybackslash\hspace{0pt}}m{#1}}
\newcolumntype{R}[1]{>{\raggedleft\let\newline\\\arraybackslash\hspace{0pt}}m{#1}}
\resizebox{0.9\columnwidth}{!}{
	\begin{tabular}[c]{C{2.5cm}|C{2.5cm}|C{3.5cm}|C{3.5cm}}
	\hline
	\hline
	Number of States & Number of Mixtures & MAP WEER (\%) & MVE WEER (\%) \\
	\hline
	\hline
	1 & 128 & 5.56 & 4.65 \\
	8 & 16 & 5.38 & 5.03 \\
	\hline
	1 & 256 & 4.81 & 4.51 \\
	16 & 16 & \textbf{4.55} & 4.44 \\
	\hline
	1 & 512 & 5.15 & 4.72 \\
	32 & 16 & 4.59 & \textbf{4.31} \\
	\hline
	1 & 1024 & 5.02 & 4.72 \\
	32 & 32 & 5.11 & 4.71 \\
	\hline
	\hline
	\multicolumn{2}{C{5cm}|}{Average} & 5.02 & 4.64 \\
	\hline
	\hline
\end{tabular}
}
 
\label{table:w1_t180}
\end{table}

\begin{table}[htbp!]
\centering
\caption{WEER performance under different HMM model configurations after VAD. The decision window duration is 1.01 seconds. The \textbf{enrollment utterance is 105 seconds long} for each speaker.} 
\newcolumntype{L}[1]{>{\raggedright\let\newline\\\arraybackslash\hspace{0pt}}m{#1}}
\newcolumntype{C}[1]{>{\centering\let\newline\\\arraybackslash\hspace{0pt}}m{#1}}
\newcolumntype{R}[1]{>{\raggedleft\let\newline\\\arraybackslash\hspace{0pt}}m{#1}}
\resizebox{0.9\columnwidth}{!}{
	\begin{tabular}[c]{C{2.5cm}|C{2.5cm}|C{3.5cm}|C{3.5cm}}
	\hline
	\hline
	Number of States & Number of Mixtures & MAP WEER (\%) & MVE WEER (\%) \\
	\hline
	\hline
	1 & 128 & 6.85 & 6.15 \\
	8 &	16 & 6.56 & 6.49 \\
	\hline
	1 & 256 & 6.40 & 5.94 \\
	16 & 16 & \textbf{6.27} & \textbf{5.76} \\
	\hline
	1 & 512 & 6.73 & 6.53 \\
	32 & 16 & 6.51 & 5.83 \\
	\hline
	1 & 1024 & 7.81 & 7.54 \\
	32 & 32 & 7.76 & 6.21 \\
	\hline
	\hline
	\multicolumn{2}{C{5cm}|}{Average} & 6.86 & 6.31 \\
	\hline
	\hline
\end{tabular}
}

\label{table:w1_t105}
\end{table}

\subsection{Performance of conventional speaker verification using
AVA real-time decisions}
\label{subsec:expr_sv}

In the scenario of conventional speaker verification, each test utterance
is assumed to include the speech of only speaker with a claimed identity.
A decision is made on the speaker identity by comparing a threshold with the
log-likelihood score of the entire test utterance given the claimed
speaker model, while in the case of AVA, a decision needs to be made on
each short-duration test window in real-time because the test utterance
may undergo change of speaker at any time instant.

Here, we are interested in exploring the performance of the window-based
modeling and testing scheme of AVA in a conventional utterance-based
speaker verification task. Under the assumption that each utterance is
spoken by a single talker, we form the verification decision for the
entire utterance by instituting a majority vote from the AVA short-time
window-based decision sequence. Specifically, for a test utterance of duration $T$, if
more than $\lfloor T/\delta_f\rfloor / 2$ of the decisions for the window-based tests
are ``true speaker (impostor)'', the final decision for this
utterance will be ``true speaker (impostor)''. With the utterance-level
ground truth, EER can be computed in the way as described in
Section \ref{sec:metric} by varying the threshold. 

With the AVA real-time decisions in Section
\ref{subsec:expr_model_config}, an EER of 0.00\% is achieved under all
HMM configurations for the utterance-based conventional speaker
verification task using the AVA database.


\subsection{Statistical validation of performance}
\label{subsec:expr_kfold}

Cross validation is an effective statistical method to test the
generalizability of a model to new or unseen data~\cite{kfold}. In the
text-independent real-time speaker verification task, in which the
data is randomly divided into roughly equal K subsets and for each
validation trial, one of the K subsets is used as the testing set
while the rest of the K-1 subsets are put together to form a enrollment
set. The enrollment set is fit to a model and predictions are made on
the testing data based on the trained model. In each round, the
validation is repeated K times and results are then averaged over the
K validation trials or folds.

To ascertain the statistical significance of the obtained performance,
we use K-fold cross validation to systematically check the accuracy of
the speaker models for unseen data. We set K=3 to keep the duration of
the enrollment set for each speaker to 240 seconds, on average, which
makes the results comparable with the evaluations in
Section~\ref{subsec:expr_model_config}. We run 5 rounds of the 3-fold
cross validation and average over the results for the folds and
rounds. The results are shown in Table~\ref{table:kfold}, in which we
give the average WEER of the rounds and the 95\% confidence range (CR)
to indicate the variation of the WEER. We notice that the CR for the MAP
algorithm are smaller than for MVE algorithm.


Furthermore, when comparing the results in Table~\ref{table:kfold}
with the results in Table~\ref{table:w1_t240}, we notice that the WEER
is less than that achieved for the same model. This is because the
enrollment and testing datasets, in contrast to the evaluations done in
the preceding section, are more matched in terms of the material,
despite being selected randomly. As is mentioned in Section
\ref{sec:expr}, the dataset for each speaker consists of four parts:
rainbow passage, 8 repeated pass-phrases, Harvard sentences and digit
pairs. For 3-fold validation, we randomly select roughly a third of
the utterances from each part, combine them to be the testing set and
use the rest two-third as the enrollment set. This means that some of
the repeated pass-phrase utterances will be shared between the
enrollment and test sets. A similar sharing may occur for the digit
pairs. Thus, a lower WEER is obtained in this case.

\begin{table}
\centering
\caption{Average WEERs with 95\% CR for the model configuration shown
  with 5 runs each of the 3-fold validation of the MAP and MVE
  algorithms on the AVA database. All results include VAD decision
  modulation. The testing window duration is set at 1.01 second.}
\newcolumntype{L}[1]{>{\raggedright\let\newline\\\arraybackslash\hspace{0pt}}m{#1}}
\newcolumntype{C}[1]{>{\centering\let\newline\\\arraybackslash\hspace{0pt}}m{#1}}
\newcolumntype{R}[1]{>{\raggedleft\let\newline\\\arraybackslash\hspace{0pt}}m{#1}}
\resizebox{0.9\columnwidth}{!}{
\begin{tabular}{C{2.5cm}|C{2.5cm}|C{3.5cm}|C{3.5cm}}
\hline
\hline
    Number of States & Number of Mixtures &Average MAP WEER $\pm$ CR(\% ) &
    Average MVE WEER $\pm$ CR (\% )\\
\hline
 \hline
1 & 512 & 2.42 $\pm$ 0.09& 2.01 $\pm$ 0.12\\
\hline
32 & 16 & 2.29 $\pm$  0.11 & 1.79 $\pm$  0.12 \\
\hline
\hline
  \end{tabular}}

  \label{table:kfold}
\end{table}


\subsection{Performance evaluation on NIST SRE} 

The NIST SRE Training and Test Sets are widely used to evaluate the
performance of speaker verification systems.  In NIST SRE, the decisions
are made on each test utterance based on the speaker models trained with
the provided training data and the performance is evaluated with respect
to the ground truth. We notice that cross-talk components exist in these datasets, i.e., even though a test utterance is labeled as coming from a certain speaker in the label, some portion of the utterance is actually from another speaker (e.g., see 2001 NIST SRE). Although these crosstalk
components may not substantially affect the performance evaluation
designed for NIST's utterance-based authentication, it does not suit the
evaluation of the real-time, window-based AVA system as the real identity
of each sliding window of the speech signal is not known.
To verify the effectiveness our method on large and publicly available
datasets, we sift out the cross-talk components in NIST SRE 2001 dataset
\cite{nist01_eval_plan} and evaluate the performance of AVA using both
i-vector and the proposed method with the remaining speech signal.

In NIST SRE 2001, there are 100 female target speakers and 74 male
target speakers with around 2 minutes of enrollment data for each speaker.
In addition, there are 2038 test segments, each of which has a duration
varying between 15 to 45 seconds.  Each test segment will be evaluated
against 11 hypothesized speakers of the same gender as the target speaker.
One of the 11 hypothesized speakers is the true speaker present in the
test segment and rest of them are impostors.

Since the cross-talk components in NIST SRE 2001 has significantly lower
energy per frame than the speech signal from the target speaker, a
decision of ``cross-talk'' is made for a speech frame if its log mel
energy is below a certain threshold. The decision for each window of
frames is then made by taking the consensus of the threshold decisions
within that window. The windows labeled as ``cross-talk'' are 
eliminated in both the enrollment and test utterances and ignored in the
experiments. After sifting, we kept about 75\% of windows in the enrollment
data and 95\% of windows in the test data.

We first evaluated the i-vector technique in both the utterance-based conventional speaker
verification task and the AVA task in exactly the same way as described in Section
\ref{sec:ava_ivector}. The WEER results with respect to the number of
mixtures are listed below. 

\begin{table}[htbp!]
\centering
\caption{WEER (\%) of AVA and EER (\%) of conventional speaker
verification (SV) using i-vector on NIST SRE 2001 dataset with different
UBM-GMM configurations. The decision window duration for AVA is 1.01
seconds.} 
\newcolumntype{L}[1]{>{\raggedright\let\newline\\\arraybackslash\hspace{0pt}}m{#1}}
\newcolumntype{C}[1]{>{\centering\let\newline\\\arraybackslash\hspace{0pt}}m{#1}}
\newcolumntype{R}[1]{>{\raggedleft\let\newline\\\arraybackslash\hspace{0pt}}m{#1}}
\resizebox{0.70\columnwidth}{!}{
	\begin{tabular}[c]{C{2.5cm}|C{3.5cm}|C{3.5cm}}
	\hline
	\hline
	 Number of Mixtures & AVA WEER (\%) & SV EER (\%) \\
	\hline
	\hline
	 64 & 28.02 & 7.51 \\
	\hline
	 128 & 26.96 & \textbf{6.67} \\
	 \hline
	 256 & \textbf{26.53} & 7.26 \\
	 \hline
	 512 & 26.74 & 7.90 \\
	 \hline
	 1024 & 27.66 & 8.44 \\
	 \hline
	 \hline
	\end{tabular}
}
\label{table:ivector_nist01}
\end{table}

We then apply the proposed training and testing method described in
Sections \ref{sec:train} and \ref{sec:test} to the AVA task on NIST SRE
2001 and obtain the WEER results below. We also compare the performance
difference between MAP and MVE training.

\begin{table}[htbp!]
\centering
\caption{WEER (\%) of AVA using MAP and MVE on
NIST SRE 2001 with different HMM model configurations. The decision window duration is 1.01 seconds.} 
\newcolumntype{L}[1]{>{\raggedright\let\newline\\\arraybackslash\hspace{0pt}}m{#1}}
\newcolumntype{C}[1]{>{\centering\let\newline\\\arraybackslash\hspace{0pt}}m{#1}}
\newcolumntype{R}[1]{>{\raggedleft\let\newline\\\arraybackslash\hspace{0pt}}m{#1}}
\resizebox{0.90\columnwidth}{!}{
	\begin{tabular}[c]{C{2.5cm}|C{2.5cm}|C{3.5cm}|C{3.5cm}}
	\hline
	\hline
	 Number of States & Number of Mixtures & MAP WEER (\%) & MVE WEER (\%) \\
	\hline
	\hline
	 1 & 128 & 24.29 & 23.67 \\
	 8 & 16 & 24.07 & 23.54 \\
	 \hline
	 1 & 256 & 24.12 & 23.55 \\
	 16 & 16 & 23.66 & 22.91 \\
	 \hline
	 1 & 512 & 24.27 & 23.34 \\
	 32 & 16 & 23.55 & \textbf{22.65} \\
	 \hline
	 1 & 1024 & 24.55 & 23.63 \\
	 32 & 32 & 23.73 & 22.77 \\
	 \hline
	 \hline
\end{tabular}
}
\label{table:map_mve_nist01}
\end{table}

By comparing Tables \ref{table:ivector_nist01} and
\ref{table:map_mve_nist01}, we see that the proposed method achieves
3.88\% absolute gain over i-vector for the AVA task when the window
duration is 1.01s. For the conditional speaker verification, our
i-vector based system achieves 6.67\% EER, which is 1.61 \% absolutely better than the
UBM-GMM baseline EER 8.28 \% reported in \cite{nist01_baseline} on NIST
2001 SRE. 

As a cross reference, for conventional speaker verification task, the
i-vector achieves 6.02\%-7.07\% and 4.77\%-5.15\% EERs for the male and
female parts, respectively, of telephone data in the core condition of
NIST SRE 2008 \cite{nist08_telephone} (the condition most similar to NIST SRE
2001); it achieves an EER of 22.01\% on NIST SRE
2008 core condition when the test utterances are truncated to 2
seconds \cite{ivector_short_2012}.

The EER performance gain over i-vector on NIST SRE justifies the
generalizability of the proposed method to the standard public datasets with large
amount of speakers for the AVA task.

\section{Conclusions}
\label{sec:conclude}

We present an ensemble of techniques that enable the active voice
authentication. The difference between AVA and traditional speaker
verification is significant: AVA makes a decision on the speaker identity
at every time instant while the latter task makes a one-time decision on
the speaker identity after the entire test utterance is obtained.
Therefore, the major challenges for the AVA task is to train accurate
speaker models using minimal amount of data for active and continuous
identity authentication with very short test signals.  

We first show that the i-vector technique is not suitable for the AVA
task since the performance degrades sharply as the duration of the test
segment becomes extremely short. In our AVA system, these challenging
requirements are satisfied by matching the training and testing condition,
adapting the SI model to the data of each individual
speaker using MAP and MVE training. We perform sequential testing with
the MVE trained model. 
In our offline evaluation of the system on the database we
recorded, the system achieves 3-4\% average WEER when the testing window
duration is just 1 second, which is far beyond human capabilities. Statistical validation is conducted via K-fold cross validation. 

From the experimental results, the WEER performance does not change too
much when the total number of mixtures goes beyond 512. We
use the model configuration with 1 state and 512 mixtures as it provides
an acceptable trade-off between the training time of the algorithm and
the WEER performance. We show that the proposed approach can be
generalized to a standard public databases with large amount of speakers
by showing that the proposed methods outperforms i-vector approach by 3.8\% absolute on NIST SRE 2001. 

We decided to use about 180 seconds of voice data to train the model
for a new user. We consider this amount of enrollment data to be acceptable 
without inducing the fatigue factor on the part of the user. It gives a 
good WEER performance at 4-5\%. 
More enrollment data will further reduce the WEER although at the expense 
of the registering user's time.
MVE provides an average performance improvement
from 0.5-1\% but requires much more training time. 
The total time for user registration is 6-10 minutes.

We select the 1.01 s window duration for the demonstration system
as it provides a reasonable WEER of approximately 5\% for the
configuration of the HMM model 
and the decision latency remains within the acceptable near real-time
requirement. Using a longer duration window provides a better
performance but produces noticeable delays in decision.

The current version of the voice authentication system assumes
operation in a low-noise environment, which means its performance will
be best in an indoor office locale. The model that the system builds
for the user can account for some minor environmental variations but
it does not take voice variability of a speaker, e.g., the Lombard
effects, into account. A change in the audio path, e.g., if an
external microphone is used, may require rebuilding the user model
and/or the SI model. More research is needed to
improve these aspects of the demonstration system. 


Currently, prevalent automatic voice assistants such as Google Home and Amazon Alexa are equipped with an authentication system at the front end to verify users' identities based on their pronunciation of a fixed wake word. 
In the future, we will explore the optimal way that an AVA can work together with the speaker verification system after a user has obtained his/her access. We will also investigate the methods to combine the AVA and the initial one-time speaker verification scores to provide more reliable continuous monitoring of real-time identities.

%
%
	

\section*{Acknowledgement}
The authors would like to thank Chao Weng and Antonio Moreno-Daniel for their help on the AVA system.

This work was supported by Defense Advanced
    Research Projects Agency (DARPA) under the Active Authentication
    program with Li Creative Technologies as the primary contractor
    and Georgia Tech the sole subcontractor. The term ``active voice
    authentication'' (AVA) was used by Li Creative Technologies in the
    title of the DARPA supported project.



\section*{References}
\bibliography{refs}
\bibliographystyle{unsrt} 


\section*{Short Biographies}

\textbf{Zhong Meng} received his Ph.D. degree from Georgia Institute of Technology in 2018. He is currently an Applied Scientist in Speech \& Language Group at Microsoft Corporation. Before joining Microsoft, he worked at AT\&T Labs Research, Mitsubishi Electric Research Labs and Microsoft Research as a Research Intern. He has been working on domain adaptation, end-to-end speech recognition, adversarial learning for robust speech recognition, speech enhancement and speaker verification, discriminative training, and multi-channel speech recognition. His adversarial domain adaptation work was nominated for the Best Paper Award at IEEE Automatic Speech Recognition and Understanding (ASRU) Workshop in 2017. 


\textbf{M. Umair Bin Altaf} is a Senior Research Scientist at Pindrop Inc.  He received a Ph.D. degree in Electrical and Computer Engineering from Georgia Institute of Technology, Atlanta, GA, USA in 2015. He has been with Avaya Labs and Broadcom Inc. as a Research Intern, working on robust speech recognition and speech coding. His research interests lie in the general area of audio signal processing with special focus on speech recognition, machine learning, and environmental sound representation.

\textbf{Biing-Hwang (Fred) Juang} is the Motorola Foundation Chair Professor and a Georgia Research Alliance Eminent Scholar at Georgia Institute of Technology. He received a Ph.D. degree from University of California, Santa Barbara and had conducted research at Bell Labs, including serving as Director of Acoustics and Speech Research. He joined Georgia Tech in 2002. Prof. Juang’s academic distinctions include: several best paper awards, IEEE Fellow, the Technical Achievement Award from the IEEE Signal Processing Society for contributions to the field of speech processing and communications, Bell Labs Fellow, member of the US National Academy of Engineering, Academician of Academia Sinica, the IEEE J.L. Flanagan Field Medal Award in Audio, Speech and Acoustics, and a Charter Fellow of the National Academy of Inventors.


\end{document}